\documentclass[apjl]{emulateapj}
\usepackage{ulem}

\shorttitle{CROWDING OUT OF GIANTS BY DWARFS}

\shortauthors{Ogihara et al.}

\begin{document}

\title{CROWDING OUT OF GIANTS BY DWARFS:
AN ORIGIN FOR THE LACK OF COMPANION PLANETS IN HOT JUPITER SYSTEMS}

\author{Masahiro Ogihara}
\affil{Nagoya University,
Furo-cho, Chikusa-ku, Nagoya, Aichi 464-8602, Japan}
\email{ogihara@nagoya-u.jp}

\author{Shu-ichiro Inutsuka}
\affil{Nagoya University,
Furo-cho, Chikusa-ku, Nagoya, Aichi 464-8602, Japan}
\and
\author{Hiroshi Kobayashi}
\affil{Nagoya University,
Furo-cho, Chikusa-ku, Nagoya, Aichi 464-8602, Japan}

\begin{abstract}
We investigate formation of close--in terrestrial planets from planetary 
embryos under the influence of a hot Jupiter (HJ) using gravitational \textit{N}--body 
simulations that include gravitational interactions between the gas disk and the 
terrestrial planet (e.g., type I migration). 
Our simulations show that several terrestrial planets efficiently form outside the orbit of 
the HJ, making a chain of
planets, and all of them gravitationally interact directly or indirectly with the HJ through resonance, which leads to inward migration of the HJ. 
We call this mechanism of induced migration of the HJ as ``crowding out.'' 
The HJ is eventually lost by collision with the central star,
and only several terrestrial planets remain. 
We also find that the efficiency of the crowding--out effect depends on model parameters;
for example, the heavier the disk is, the more efficient the crowding out is.
When planet formation occurs in a massive disk, the HJ can be lost to the 
central star and is never observed.
On the other hand, for a less massive disk, the HJ and terrestrial planets can 
coexist; however, the companion planets can be below the detection limit of current
observations. In both cases, systems with the HJ and terrestrial planets have little chance
for detection. Therefore, our model naturally explains the lack of companion planets in HJ 
systems regardless of the disk mass. In effect, our model provide a theoretical prediction
for future observations; additional planets can be discovered just outside the HJ, and their
masses should generally be small.
\end{abstract}

\section{INTRODUCTION}
\label{sec:intro}
Recent observations have found multiple--planet systems whose architectures differ 
strikingly from that of our solar system. One of the most distinctive features is a lack of 
companion planets in HJ systems (e.g., \citealt{latham_etal11}; \citealt{steffen_etal12}).
Although some exceptions are being brought out by the \textit{Kepler} survey, this trend 
is clearly seen in recent observational results.
Several ideas for the inhibition of planet formation in the vicinity of HJs
were presented (\citealt{fogg_nelson07}; \citealt{raymond_etal11}).
We propose a new way of explaining this observational property.

The combination of radial velocity and transit detection observations provides information on the
average densities of planets (e.g., \citealt{borucki_etal10}), which shows that a significant fraction
of HJs are gaseous giant planets.
This clearly means that HJs are surrounded by gaseous components in protoplanetary disks, 
at least during their formation stages.
A natural question in this regard is what happened in the protoplanetary disks in the 
presence of HJs. It might be interesting to theoretically probe the condition of planet formation
in the presence gaseous giants.

We investigate the formation of terrestrial planets in the presence of an HJ.
For the origin of HJs, although type II migration (e.g., \citealt{lin_etal96}) and tidal 
circularization of high--eccentricity planets 
(e.g., \citealt{ford_rasio06}; \citealt{nagasawa_etal08}) were proposed, we here
introduce a
hybrid scenario in which HJs can be formed through gravitational instability prior to growth
of terrestrial planets \citep{inutsuka09}.
Recent progress in our understanding of protostar formation provides an explanation regarding 
the formation of protoplanetary disks and their accretion onto the central protostar in their early 
evolutionary phases (see review by \citet{inutsuka12}). One of the findings from high--resolution 
numerical simulations that 
describe the gravitational collapse of molecular cloud cores indicates that new--born 
protoplanetary disks are gravitationally unstable and produce multiple planetary mass objects 
that tend to chaotically migrate in the massive disks (\citealt{vorobyov_basu10}; \citealt{machida_etal11}). 
Although the final fates of those 
planetary mass objects are not entirely clear, the objects that were formed relatively late may 
survive and remain as gaseous giants in various locations in the disks. Therefore it is natural to 
consider the planet formation process under the influence of 
gaseous giants in protoplanetary disks.

The letter proceeds as follows. In Section~\ref{sec:model}, we describe
our numerical model; in Section~\ref{sec:results}, we present the results of simulations.
In Section~\ref{sec:condition}, we derive the condition for migration of HJ. In 
Section~\ref{sec:discussion}, we present a discussion.

\section{MODEL DESCRIPTION}
\label{sec:model}
An HJ with Jovian mass 
is initially placed at $a_{\rm HJ} = 0.05 {\rm AU}$ on a circular orbit, 
where $a_{\rm HJ}$ is the semimajor axis
of the HJ. Under the influence of the HJ, formation of solid planets from planetary
embryos are investigated using numerical simulations that combine the \textit{N}--body code
and the semianalytical code.

The simulation domain of \textit{N}--body calculations is set between 0.02 and 0.5 AU
and the gravitational forces between all the bodies are calculated (in the same manner
as in previous works by \citealt{ogihara_etal07}; \citealt{ogihara_ida09,ogihara_ida12}).
For the initial solid disk, the solid surface density is assumed as
\begin{eqnarray}
\Sigma_{\rm d} =10 f_{\rm d}\left(\frac{r}{1~{\rm AU}}\right)^{-3/2}{\rm ~g~cm^{-2}},
\label{eq:surface_solid}
\end{eqnarray}
where $f_{\rm d}$ is a scaling factor 
and $r$ is the radial distance from the central star. 
According to this density distribution, planetary embryos are placed with
the isolation mass of $M_{\rm iso} = 2 \pi a \Delta a \Sigma_{\rm d}$,
where $\Delta a$ is the width of the feeding zone and assumed to be 
$\simeq10$ Hill radii. 
Note that protoplanets reach the isolation masses for the parameters we are
concerned with, unless significant loss of surface density caused by collisional
fragmentation occurs \citep{kobayashi_etal10}.

The effect of type I migration is considered; thus, protoplanets 
migrate in to the simulation domain from beyond 0.5 AU. We also simulate the growth and 
migration of protoplanets
beyond 0.5 AU not by \textit{N}--body code but by semianalytical code in the same
manner as in the population synthesis model (e.g., \citealt{ida_lin08}), where protoplanets
migrate after their growing. When protoplanets reach the boundary $(a=0.5 {\rm AU})$,
the bodies are added to the \textit{N}--body code.

We incorporate the effects of eccentricity and inclination damping and 
orbital migration due to gravitational interaction with the gas disk 
(e.g., \citealt{ogihara_ida09}; \citealt{ogihara_etal10}). 
The decay rate of semimajor axis is \citep{tanaka_etal02}
\begin{eqnarray}
t_a &=& \frac{1}{C_{\rm I}}\frac{1}{2.7+1.1q(r)} \left(\frac{M}{M_*}\right)^{-1}\left(\frac{\Sigma_{\rm g} r^2}{M_*}\right)^{-1}\left(\frac{c_{\rm s}}{v_{\rm K}}\right)^{2} \Omega^{-1},\\
&=& 1.6 \times 10^3 C_{\rm I}^{-1} f_{\rm g}^{-1}\left(\frac{2.7+1.1q(r)}{4.35}\right)^{-1}\left(\frac{r}{0.1~{\rm AU}}\right)^{3/2}\nonumber\\&   & \times 
\left(\frac{M}{M_\oplus}\right)^{-1}\left(\frac{M_*}{M_\odot}\right)^{-1/2}\left(\frac{L_*}{L_\odot}\right)^{1/4}{\rm ~yr},
\label{eq:a-damp}
\end{eqnarray}
where $C_{\rm I}$ and $-q(r)$ denote the type I migration efficiency factor 
\citep{ida_lin08} and
the gas surface density gradient ($q(r) = - d\ln \Sigma_{\rm g}/d\ln r$), respectively.\footnote{Note that the direction and rate of the type I migration can be altered in a 
non--isothermal disk; several factors that depend on the local
gradient of temperature and entropy should be added to the above equations
(e.g, \citealt{paardekooper_etal10}).}
The migration can be halted when the planets are captured in a mean motion resonance
with the HJ (e.g., \citealt{ogihara_kobayashi13})
or they reach a region with positive density gradient. The latter corresponds 
to locations on the disk inner edge (e.g., magnetospheric cavity) or the outer edge of
a gap that is opened by the HJ; terrestrial planets in these locations can gain a positive
torque from the disk by the steep surface density gradient \citep{masset_etal06}.
Planets stop their migration at density gaps or resonance locations, whichever are first
encountered by the planets (i.e., whichever is at the more distant location from the star).

\begin{figure}[htbp]
\epsscale{0.8}
\plotone{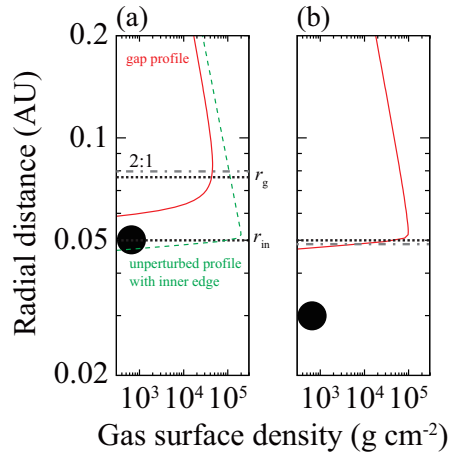}
\caption{The dashed line represents the unperturbed disk profile for $f_{\rm g}=1$
with the inner edge, where the gas density smoothly vanishes at 0.05 AU
with the width of the disk scale height.
(a) The solid line shows the disk profile, in which a density gap is opened by the HJ
at 0.05 AU. 
The dotted lines indicate the location of $q(r) = -2.7/1.1$ for this gap profile,
which exist inside the 2:1 resonance with the HJ (dot--dashed line).
(b) The disk profile for the HJ at 0.03 AU.}
\label{fig:model}
\end{figure}

The unperturbed gas surface density is considered to be of the form
\begin{equation}
\label{eq:surface_gas}
\Sigma_{\rm g} = 2400 f_{\rm g} \left(\frac{r}{1~{\rm AU}}\right)^{-3/2}\exp\left(\frac{-t}{t_{\rm dep}}\right){\rm ~g~cm^{-2}},
\end{equation}
where $f_{\rm g}$ is a scaling factor \citep{hayashi81}. We usually assume $f_{\rm g}=1$ in our simulations. 
For the case of $f_{\rm d}/f_{\rm g}=1$, the solar metallicity is considered. 
Note that there is a vast range in the metallicity of stars; we vary the values of
$f_{\rm d}$ and/or $f_{\rm g}$ in order to change the metallicity.
The gas disk exponentially dissipates on the depletion timescale of $t_{\rm dep} 
=10^6 {\rm yr} \simeq 3 \times 10^7 T_{\rm K}$, where $T_{\rm K}$ is the orbital period at 0.1 AU. 
The unperturbed gas surface density profile is shown by the
dashed line in Figure~\ref{fig:model}(a). 
We consider a disk inner edge, $r_{\rm in}$, at 0.05 AU from the central star
and inside of which there is a cavity. 
The HJ opens up a gap near its orbit; for gap profile computations we use an analytical
description developed by \citet{crida_etal06},
which depends on the
disk viscosity. In this study, relatively high value for the turbulent viscosity is
assumed $(\alpha = 10^{-2})$, where $\alpha$ denotes the disk viscosity.
We will discuss the dependence of the disk viscosity in the following paper (Ogihara et al., in prep.).
The gas surface density profile, which includes the density
gap, is shown by the solid line in Figure~\ref{fig:model}(a). 
At the gap edge, $q(r)$ becomes
smaller than $-2.7/1.1$, which cause a positive torque on the planet.
The location of $q(r) = -2.7/1.1$ for the gap edge $r_{\rm g}$ is shown by the
dotted line in Figure~\ref{fig:model}.
The location of the 2:1 resonance with the HJ is shown by the dot--dashed line,
which is further from the central star than the gap assuming $a_{\rm HJ} = 0.05 {\rm AU}$.
When the HJ comes very close to the central star $(\lesssim 0.03 {\rm AU})$, the locations of
the 2:1 resonance and the gap edge move inside $r_{\rm in}$ (Figure~\ref{fig:model}(b)).

If HJs move to very close--in orbits, the tidal interaction with the star would be important.
The migration timescale induced by the tidal torque is \citep{goldreich_soter66}
\begin{eqnarray}
t_{a,{\rm tide}} &=& \left|\frac{a}{\dot{a}}\right|
=\frac{Q_*}{3 k_{2,*}}\left(\frac{M}{M_*}\right)^{-1}\left(\frac{a}{R_*}\right)^{5} \Omega^{-1},
\end{eqnarray}
where $Q_*$ and $k_{2,*}$ are the tidal dissipation function and the love number of the 
central star, respectively. Assuming $M_* = M_\odot$ and $R_* = R_\odot$, the
timescale is reduced to
\begin{eqnarray}
t_{a,{\rm tide}} &=& 6\times 10^{9}\left(\frac{Q_*}{10^6}\right)\left(\frac{k_{2,*}}{0.3}\right)^{-1}\left(\frac{M_{\rm HJ}}{M_{\rm J}}\right)^{-1}\left(\frac{a}{0.03 {\rm AU}}\right)^{6.5} {\rm yr},
\label{eq:t_a_tide}
\end{eqnarray}
which is a strong function of $a$. We also incorporate
this damping in our simulations, adopting a slightly smaller valuer of $Q_* = 10^5$
to reduce the computational cost, which does not change our conclusions for
$Q_* \lesssim 10^{6}$.
Note that if HJs reach 0.02 AU, they
can fall onto the star within several $10^8$ yr even assuming conservative values of $Q_*=10^6$ and
$R_* = R_\odot$. Thus, we set the inner boundary of the computational region at 0.02 AU.

\section{RESULTS}
\label{sec:results}
\begin{figure}[htbp]
\epsscale{0.9}
\plotone{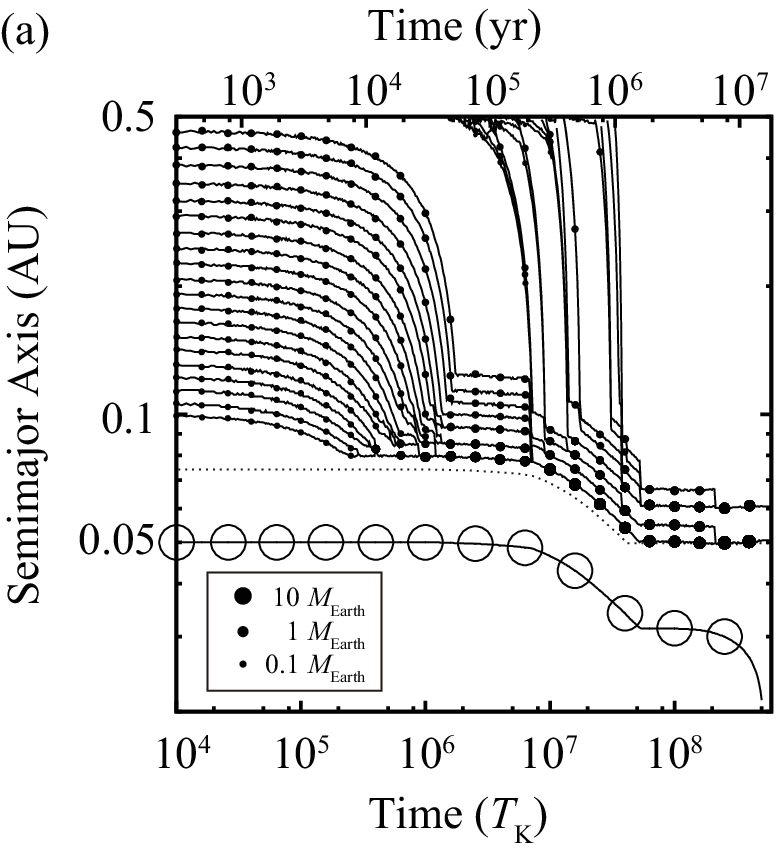}
\plotone{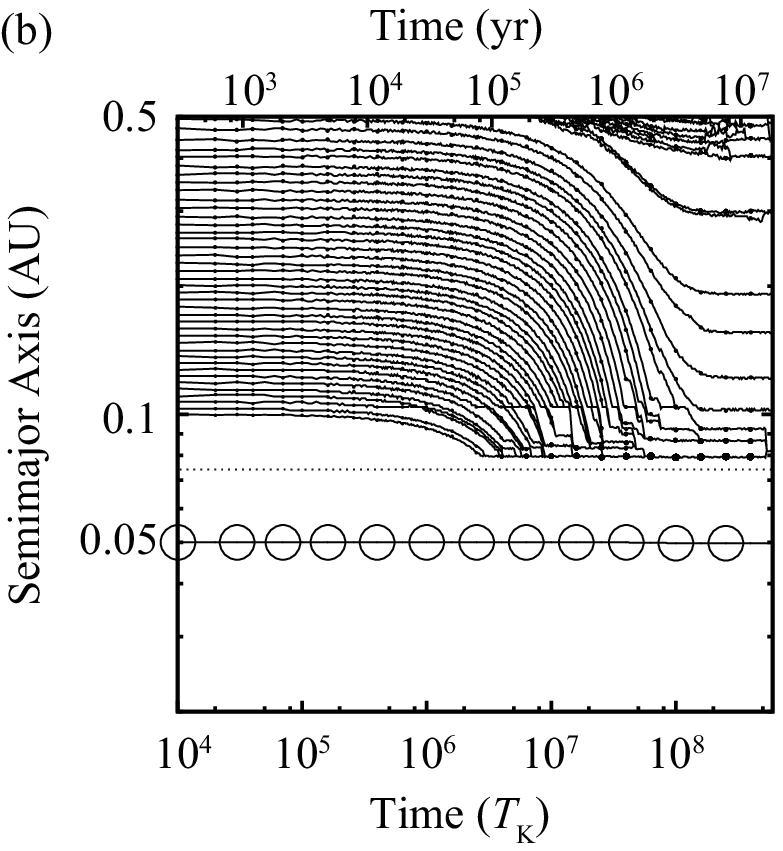}
\plotone{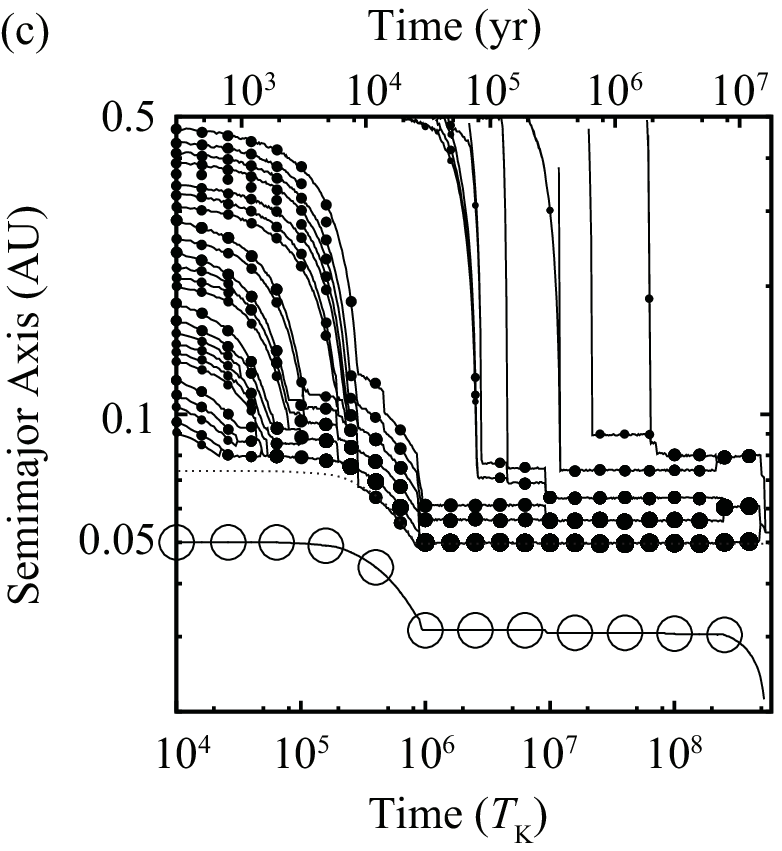}
\caption{Evolution of systems for (a) the fiducial model, (b) the decreased
solid surface density $(f_{\rm d}=0.1)$, and (c) the increased solid surface density
$(f_{\rm d}=10)$. The open circles connected with solid lines represent the HJ, whereas
the solid circles show the evolution of other bodies. The dotted lines indicate the location of 
$q(r) = -2.7/1.1$ in the gas disk.}
\label{fig:t_a}
\end{figure}

In Figure~\ref{fig:t_a}(a), we show the time evolution of semimajor axis for our fiducial
model parameters, where $f_{{\rm d}} = f_{{\rm g}} = C_{{\rm I}} = 1$ is assumed. 
The time unit is expressed by the orbital period at 0.1 AU (lower axis) and by
yr (upper axis).
All bodies are plotted as filled circles connected with solid lines, whereas the HJ is
represented by open circles. 
The dotted line represents the location of $q(r)=-2.7/1.1$; the inner disk
profile is determined by the gap opened by the HJ $(t \lesssim 9 \times 10^7 T_{\rm K})$,
whereas it is determined by the inner cavity in the late stage of the simulation.
Planetary embryos that are initially placed inside 0.5 AU undergo
inward migration. The innermost planet stops its migration when it is captured 
into the 2:1 mean motion resonance with the HJ at $t \simeq 2 \times 10^{5} T_{\rm K}$.
Other migrating protoplanets approach the inner planet, which
is captured in the 2:1 resonance, and interact with each other, leading to mergers or
captures into mean motion resonances. As a result, several planets relax to a quasi--steady
state captured in mutual mean
motion resonances, which is called a chain of resonant planets or a ``resonant chain.''

After that, several planetary
embryos whose growth is calculated using the semianalytical code migrate inward from
outside 0.5 AU.\footnote{We see an unphysical time gap between $10^6$ and $10^7 T_{\rm K}$, which can be
attributed to the fact that the growth of protoplanets inside 0.5 AU is not simulated 
in this work. Note, however, that this produces no systematic change in the final state
of planetary architecture, which is confirmed by high--resolution simulations
(Ogihara et al., in prep.).}
These protoplanets also interact with the planets in the resonant chain, and
eventually participate in the resonant chain.
Because the innermost planet is located outside the gap edge,
the planets in the resonant chain reside in the region of negative density gradient
in the gas disk and keep losing the orbital angular momentum by 
the negative type I migration torque. Through gravitational interaction between the HJ
and the planets in the resonant chain,
the HJ migrate inward by being pushed in by the resonant chain as a whole,
although the HJ itself is in the cavity of the gas disk.
We call this phenomenon a ``crowding out'' effect.
Note that the crowding out may also occur due to the interaction
between an HJ and a massive swarm of planetesimals 
feeling strong aerodynamical gas drag (\citealt{raymond_etal06}; 
\citealt{mandell_etal07}).

When the HJ comes to a very close--in orbit $(a \simeq 0.03 {\rm AU})$, 
the location of the 2:1 resonance with the HJ is comparable to the location
of the disk inner edge at 0.05 AU (Figure~\ref{fig:model}(b)). The
innermost planet in the resonant chain reaches the disk inner edge and gains the 
positive torque from the disk, which
compensates for the net negative torque exerted on the planets in the resonant chain;
the inward migration of the HJ due to the crowding--out effect is halted
at $t \simeq 9 \times 10^7 T_{\rm K}$.\footnote{Note that if $r_{\rm in}$ is closer to the star than that
assumed in this letter, the HJ migrates inside 0.03 AU due to the crowding out.}

The disk gas is depleted on the timescale of $10^{6} {\rm yr}
\simeq 3 \times 10^{7} T_{\rm K}$. Then, only the HJ moves inward because of the tidal torque from the host star,
which results in widening separation between the HJ and the innermost planet.
This leads to a deviation from the 2:1 resonance, and subsequently
some planets exhibit orbit crossing.
In this run, two giant impact events between planets are observed at 
$t \simeq 2.1 \times 10^{8} T_{\rm K}$,
which results in losses of commensurate relationships. The HJ is lost to the
central star\footnote{The HJ at 0.03 AU can fall onto the star within a few billion yeras for $Q_{*}<10^{6}$.},
and two terrestrial planets remain at the end of simulation. 
The largest planet has $M = 2.3 M_\oplus$.

Figures~\ref{fig:t_a}(b) and (c) show the results with decreased $(f_{\rm d}=0.1)$
and increased $(f_{\rm d}=10)$ solid surface density, respectively. 
Although the HJ gravitationally interacts with
the resonant chain, the inward migration of the HJ caused by the crowding out is not
seen in the result of decreased solid density $(f_{\rm d}=0.1)$ because small
planets formed in the less massive disk, feeling weak type I migration torques, cannot
effectively reduce the angular momentum of the HJ through resonant interactions.
The largest planet has a mass of $0.15 M_\oplus$ at the end of simulation.
The migration of the HJ due to the tidal torque from the star is also not observed because 
the HJ has a larger semimajor axis $(a\simeq0.05 {\rm AU})$.
Several planets have commensurate relations in the final state.
On the other hand, the crowding out of the HJ is clearly
observed in the result of increased density $(f_{\rm d}=10)$. In fact, the timescale
of inward migration of the HJ is shortened. Relatively
large planets with maximum mass of $24 M_\oplus$ remain in the final state.

\section{CONDITION FOR CROWDING OUT OF HJ}
\label{sec:condition}
We now derive the condition for inward migration of the HJ due to the crowding out
within the disk lifetime. When the HJ moves close to the central star, it can subsequently be
lost to the star because of the tidal evolution. Therefore, the condition for crowding out can
simply corresponds to the condition for the loss of HJ.

The timescale of the HJ's induced migration due to the crowding--out effect, $t_{a, {\rm HJ}}$,
which depends on the total mass in planets in the resonant chain $(M_{\rm chain})$ and 
on the migration timescale of planets in the chain $(t_{a,{\rm chain}})$, is given by
\begin{eqnarray}
\label{eq:ta_hj}
t_{a, {\rm HJ}} = -\frac{J_{\rm HJ}}{2 T_{\rm chain}}\simeq t_{a,{\rm chain}} \left(\frac{a_{\rm HJ}}{a_{\rm chain}}\right)^{1/2}\frac{M_{\rm HJ}}{M_{\rm chain}},
\end{eqnarray}
where $J_{\rm HJ}$ and $T_{\rm chain}$ are, respectively, the angular momentum of the HJ
and the total migration torque exerted on the planets in the resonant chain. 
$a_{\rm chain}$ and $t_{a,{\rm chain}}$ are the typical values for the semimajor axis
and the migration timescale of the planets in the resonant chain.

Assuming that all the bodies that migrate to near 0.1 AU are included in the resonant chain,
we derive the total mass of the migrating planets $M_{\rm chain}$. Because the type I migration speed
increases with increasing mass, bodies grow to a certain mass and then start migration.
The mass of migrating planets is expressed by the critical mass for migration,
which is determined by a balance between the accretion timescale and the migration
timescale, or the isolation mass, whichever is smaller. The isolation mass
is smaller in the inner region; thus, we can define a boundary $a_{\rm trans}$, inside
which all the migrating planets are with the isolation mass.

The total mass of the migrating planets is given by
\begin{eqnarray}
\label{eq:m_chain}
M_{\rm chain}\simeq\int^{a_{\rm max}}_{a_{\rm in}}\Sigma_{\rm d} 2\pi r dr,
\end{eqnarray}
where $a_{\rm in}$ is the radius of the inner edge of the initial solid disk, and
$a_{\rm max}$ denotes the maximum semimajor axis, inside which
solid bodies can start migration before the disk gas is depleted.
We assume $a_{\rm in} =0$ for simplicity.
When input parameters ($f_{\rm d}$, $C_{\rm I}$, and $f_{\rm g}$) are specified,
a unique value for $a_{\rm max}$ is determined.

We first consider the case where planets with the isolation mass can only migrate 
inward;
the condition for $a_{\rm trans} > a_{\rm max}$ is given by
\begin{eqnarray}
2 \left(\frac{t_{\rm dep}}{10^6 {\rm yr}}\right)^{4/3}
f_{\rm d}^{27/11} C_{\rm I}^{59/33} f_{\rm g}^{53/33}\la 1,
\end{eqnarray}
where $M_* = M_\odot$ and $L_* = L_\odot$ are assumed hereafter.
Then the maximum semimajor axis $a_{\rm max}$ is deduced
by equating the migration timescale $t_a|_{M=M_{\rm iso}}$ with the disk depletion timescale
$t_{\rm dep}$. Substituting $a_{\rm max}$ into Equation~(\ref{eq:m_chain}), we obtain
\begin{eqnarray}
\label{eq:m_chain2}
M_{\rm chain, iso} &\simeq& 0.1\left(\frac{t_{\rm dep}}{10^6 {\rm yr}}\right)^{2/3}\left(\frac{f_{\rm d}}{0.1}\right)^{3} C_{\rm I}^{2/3} f_{\rm g}^{2/3} M_\oplus.
\end{eqnarray}
Note that this should be overestimated by about a factor of 2--5 because the integral in
Equation~(\ref{eq:m_chain}) is
performed from the central star for simplicity.
Then the migration timescale of the HJ is approximated by
\begin{eqnarray}
t_{a, {\rm HJ, iso}} &\simeq& 4\times 10^7\zeta^{-1} \left(\frac{a_{\rm HJ}}{0.05 {\rm AU}}\right)^{1/2}\left(\frac{M_{\rm HJ}}{M_{\rm J}}\right)\left(\frac{t_{\rm dep}}{10^6 {\rm yr}}\right)^{-4/3}\nonumber\\&&\times 
\left(\frac{f_{\rm d}}{0.1}\right)^{-4} C_{\rm I}^{-7/3} f_{\rm g}^{-7/3}\left(\frac{a_{\rm chain}}{0.1 {\rm AU}}\right){\rm yr},
\label{eq:ta_hj_iso}
\end{eqnarray}
where a correction factor $\zeta$ is introduced. In addition to the overestimate
of $M_{\rm chain}$, the gas density profile also accounts for this correction.
Although we adopt Equation~(\ref{eq:surface_gas}) and derive the migration timescale,
both $\Sigma_{\rm g}$ and $q(r)$ become smaller around $r = 0.1 {\rm AU}$. 
We put all of these corrections into the factor $\zeta$.

When the planets migrate inward from the
outer region where the mass is determined by the critical mass for migration
$(a_{\rm trans} < a_{\rm max})$, the total mass in the resonant chain is estimated by
\begin{eqnarray}
\label{eq:m_chain1}
M_{\rm chain, crit} &\simeq& 5.6\left(\frac{t_{\rm dep}}{10^6 {\rm yr}}\right)^{5/24}
f_{\rm d}^{37/32} C_{\rm I}^{5/96} f_{\rm g}^{11/96} M_\oplus.
\end{eqnarray}
Note again that this is somewhat overestimated.
The migration timescale of the HJ is given by
\begin{eqnarray}
t_{a, {\rm HJ, crit}} &\simeq& 1\times 10^4\zeta^{-1}\left(\frac{a_{\rm HJ}}{0.05 {\rm AU}}\right)^{1/2}\left(\frac{M_{\rm HJ}}{M_{\rm J}}\right)\left(\frac{t_{\rm dep}}{10^6 {\rm yr}}\right)^{-5/12}\nonumber\\
&&\times f_{\rm d}^{-37/16} C_{\rm I}^{-53/48} f_{\rm g}^{-59/48}\left(\frac{a_{\rm chain}}{0.1 {\rm AU}}\right){\rm yr},
\label{eq:ta_hj_crit}
\end{eqnarray}
where $\zeta$ is the correction factor as mentioned above.
By comparing this expression with the actual migration timescale observed in simulations, 
we find $\zeta \simeq 0.01$.

\begin{figure}[htbp]
\epsscale{0.76}
\plotone{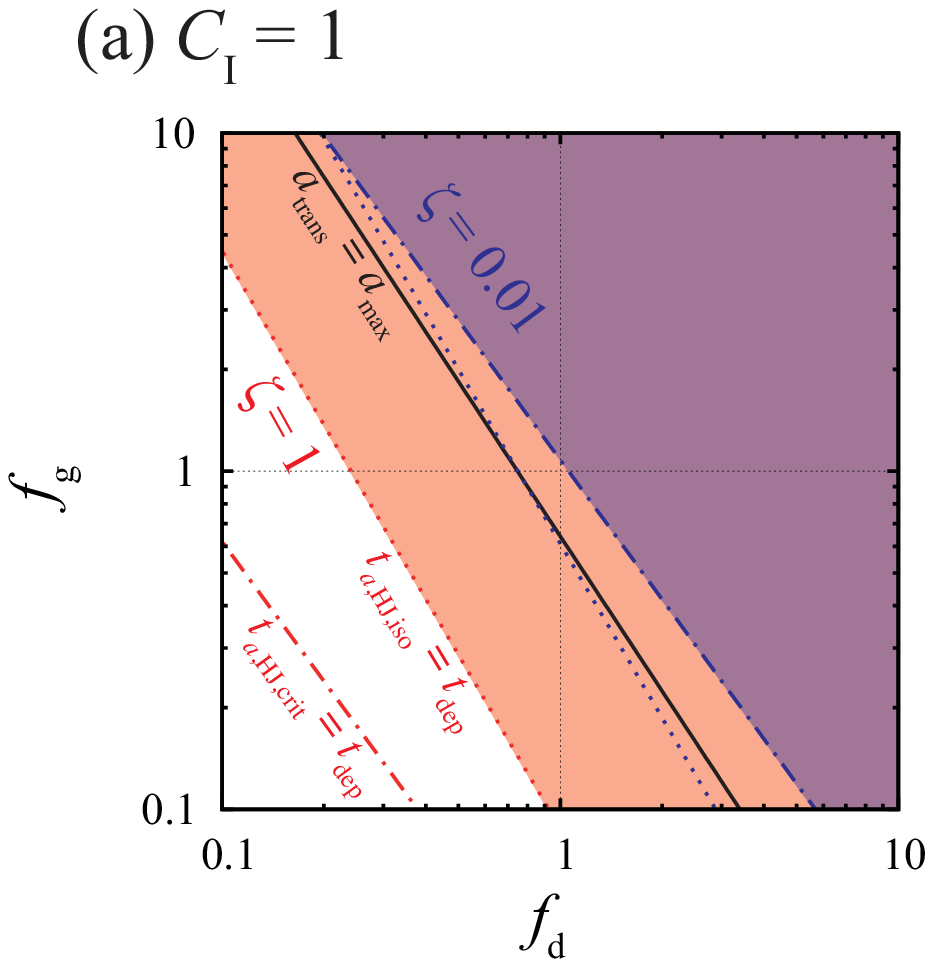}
\plotone{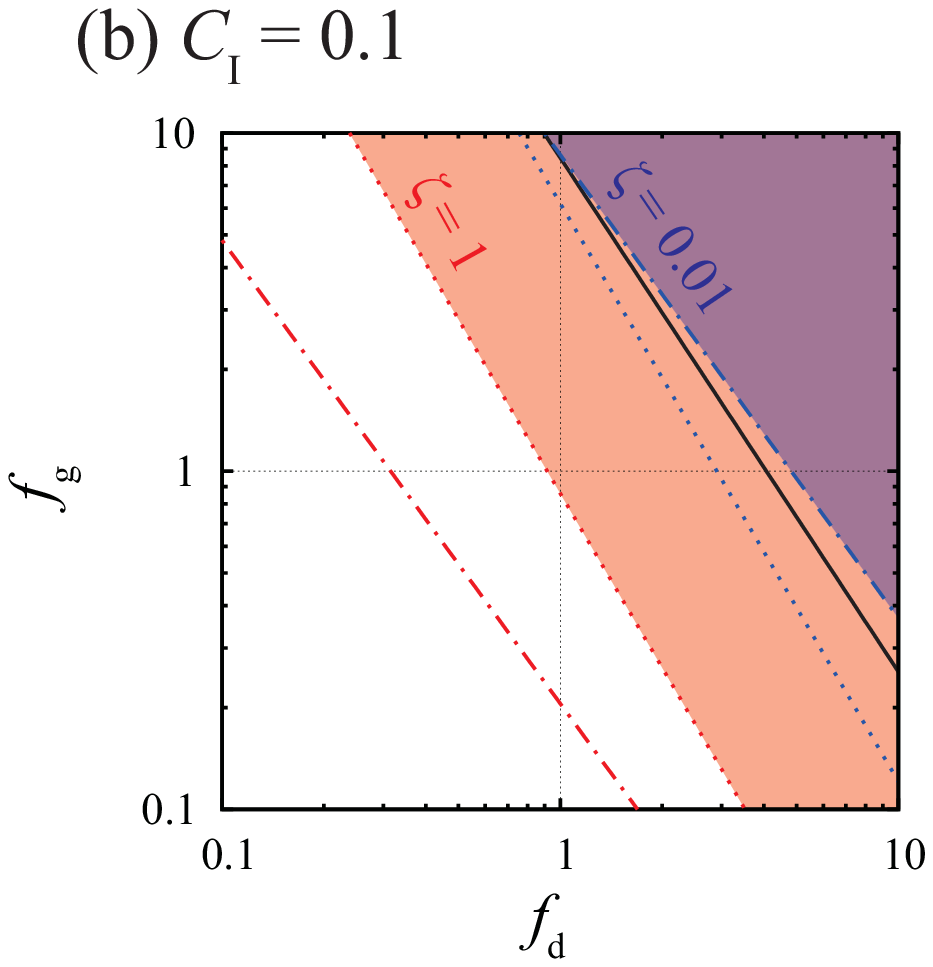}
\caption{The migration timescale of the HJ $(t_{a,{\rm HJ}})$ is compared with the disk
depletion timescale $(t_{\rm dep})$. The solid, dotted, and dash--dotted lines indicate
$a_{\rm trans}=a_{\rm max}$, $t_{a,{\rm HJ, iso}} = t_{\rm dep}$, and 
$t_{a,{\rm HJ, crit}} = t_{\rm dep}$, respectively. The regions where the crowding out
occurs are hatched; the gray--hatched (red in the online journal), and the dark--hatched
(blue in the online journal) regions show
the conditions for $\zeta=1$ and $\zeta=0.01$, respectively. (a) The case of 
$C_{\rm I} =1$. (b) The case of $C_{\rm I}=0.1$.}
\label{fig:condition}
\end{figure}

In Figure~\ref{fig:condition}, we plot the condition for migration 
$(t_{a, {\rm HJ}} < t_{\rm dep})$ due to the crowding 
out on the $f_{\rm d}-f_{\rm g}$ plane for $\zeta=1$ (gray--hatched region)
and 0.01 (dark--hatched region). Figure~\ref{fig:condition}(a) is the case where
the full type I migration rate is considered $(C_{\rm I}=1)$, whereas Figure~\ref{fig:condition}(b)
shows the condition for $C_{\rm I}=0.1$.
We find that the parametric region in which the crowding out is effective is not so limited;
in fact, the effect becomes rather common when the gas or solid surface densities are larger
than the minimum mass solar nebular model ($f_{\rm g} f_{\rm d}>1$).
The condition for crowding out is not expected to depend on the metallicity 
$(f_{\rm d}/f_{\rm g})$. There is a correlation between the stellar metallicity and the 
probability of an HJ (e.g., \citealt{fischer_valenti05}), which can be explained if
HJs are more likely to form around high--metallicity stars.

\section{POSSIBLE EXPLANATION FOR PROPERTY OF CLOSE--IN EXOPLANETS}
\label{sec:discussion}
The origin of the lack of companion planets close to HJs can be naturally explained using 
our model. When the crowding out of HJ is effective (e.g., the disk mass 
$(f_{\rm d}f_{\rm g})$ is large), the HJ can be pushed inward 
by a chain of resonant planets, leading to a collision with the central star. Therefore, the HJ 
is never observed, whereas other terrestrial planets remain around 0.1 AU. On the other 
hand, when the crowding out is not effective (e.g., in a low--mass disk), the HJ and 
companion planets may coexist. However, their masses and/or sizes are below the detection 
limit of the current survey. In both cases, systems with the HJ and close--in terrestrial planets 
have little chance to be detected.

Our model provides a theoretical prediction for future observations.
That is, if additional planets will be found just outside the HJ, the masses of planets
are likely to be small. In those systems,
the maximum mass for solid planets can be estimated by equating $t_{a,{\rm HJ}}$ 
and $t_{\rm dep}$ as
\begin{eqnarray}
M_{\rm chain, max} &\simeq& 0.6\zeta^{-1/2}\left(\frac{a_{\rm HJ}}{0.05 {\rm AU}}\right)^{1/4}\left(\frac{M_{\rm HJ}}{M_{\rm J}}\right)^{1/2}\left(\frac{t_{\rm dep}}{10^6 {\rm yr}}\right)^{-1/2}\nonumber\\
&&\times C_{\rm I}^{-1/2} f_{\rm g}^{-1/2}\left(\frac{a_{\rm chain}}{0.1 {\rm AU}}\right)^{1/2}
M_\oplus.
\end{eqnarray}
In addition, if many exoplanets that contradict our prediction are detected, it means
that some of the assumptions made in our model should be invalid. 
One possibility is that the inner edge of the gaseous disk is much larger than our choice
at the formation stage of terrestrial planets, which makes the crowding out inefficient.
Whatever observational results we obtain, the implications of our results may impose
some constraints on the planet formation theory.

\subsection*{ACKNOWLEDGMENT}
We thank the anonymous referee for useful comments.
We also thank Shoichi Oshino, Eiichiro Kokubo, and Yasunori Hori for fruitful discussions.
Numerical computations were in part conducted on 
the general--purpose PC farm at CfCA of NAOJ.

{}

\end{document}